\documentclass{sig-alternate}

\usepackage{epsfig}
\usepackage{algorithmic}
\usepackage{algorithm}
\usepackage{colortbl}
\usepackage{hhline}
\usepackage{multirow}

\begin{document}

\title{Link Based Session Reconstruction: Finding All Maximal Paths}
%
%
%
%
%

\numberofauthors{2} 
%
\author{
%
%
\alignauthor
Murat Ali Bayir\titlenote{Murat Ali Bayir is currently with Google, NY, USA, his contact email is 
bayir@google.com}\\
       \affaddr{University at Buffalo, SUNY}
\alignauthor
Ismail Hakki Toroslu\\
       \affaddr{Middle East Technical University}
}

\maketitle
\begin{abstract}

This paper introduces a new method for the session construction problem,
which is the first main step of the web usage mining process. Through experiments, it is
shown that when our new technique is used, it outperforms previous approaches in web usage mining
applications such as next-page prediction.

\end{abstract}

\keywords{Web Usage Mining, Graph Theory} 

\section{Introduction}

The purpose of Web Usage Mining (WUM)~\cite{Liu11} is to find interesting knowledge about navigation behaviors of web users. The first step of WUM includes the session construction from user logs which directly affects the quality of patterns discovered in WUM process. Previous approaches~\cite{Mobasher07} for session reconstruction has two problems. They are either using time information without link data or add artificial backward movements to complete paths in web topology which generates a noise in page view sequences. These problems can be handled by using cookies and adding client specific information in server requests. However, for various reasons, such as security and changes in the internal structure of web site, some site owners may not want to use proactive approaches at all. Instead of that, these site owners prefer to process only their raw server logs.

Our Previous method Smart-SRA~\cite{Bayir09,Bayir12} solved most of the problems mentioned above. However, it still can not capture particular user behaviors due to its greedy nature when user navigation is more complex. To overcome the problems of Smart-SRA and previous approaches, we propose a new link based technique, called as Complete Session Reconstruction Algorithm (C-SRA). C-SRA is very powerful algorithm which produces complete set of maximal paths that can be obtained from given page request sequence and web topology.

\section{Complete-SRA}

Complete Session Reconstruction Algorithm (CSRA) is a two phased session reconstruction algorithm which produces link based sessions with all possible maximal sequences. In the first phase of C-SRA, user log sequences from server logs including $<$IP,URL,Time$>$ tuples are partitioned into smaller candidate sessions such that each one of these candidate sessions satisfy both page stay and session duration time constraints mentioned in~\cite{Bayir09}. The second phase of C-SRA constructs all maximal navigation sequences from the candidate sessions generated at the first phase of the algorithm. We define session reconstruction problem as a graph problem which is called Maximal Paths in a Vertex Sequence (MPVS) as follows:

\textbf{Problem [Maximal Paths in a Vertex Sequence]:} Given vertex sequence and directed graph, determine all maximal paths in the given ordered vertex sequence.

\textbf{Input:} A possibly cyclic directed graph G = (V, E) such that $V=\{v_{1}, v_{2},$ $\ldots,$ $v_{n}\}$ is vertex set and E   $\subset$ V$\times$V is a set of edges, and a sequence of vertices $S=[vs_{1}, vs_{2},\ldots, vs_{k}]$ where each $vs_{i}$ $\in$  V (without any repetition for our problem, since the second request of the same page is always provided by the browser cache for limited time interval).

\textbf{Output:} Set $\sum$ of $\sum_{j}s$, where, each $\sum_{j} =$ $<vs_{j1},$ $vs_{j2},$ $\ldots,$ $vs_{jm}>$ is a maximal navigation sequences of S corresponding to a paths in G. That is, for every pair of consecutive vertices in a sequence $\sum_{j}$, such as $vs_{jp}$ and $vs_{j(p+1)}$, there exists an edge $(vs_{j(p)}, vs_{j(p+1)})$ $\in$ E. In addition, in order to satisfy the maximality property, there is no other sequence $\sum_{q}$ of S in $\sum$ such that $\sum_{j}$ is a sub-string of $\sum_{q}$.

Below we describe the details of C-SRA. The main part of the second phase of C-SRA corresponds to the maximal paths in a vertex sequence problem.  As an input to our algorithm, we were given user web page request sequence as vertex sequences of the web site graph, and the web site topology where vertices represent web pages and edges represent links among web pages. 

\begin{center}
\textbf{Phases of CSRA}
\end{center}

\textbf{Input:} Page request sequence of a user, given in timestamp order (UserRequestSequence) and topology of web site in adjacency matrix form (Link).

\textbf{Output:} The set of all maximal navigation sequences (MSeqSet).

\textbf{Phase 1:} Construct the candidate sessions set (CandidateSessionSet) from user page request sequence (UserRequestSequence), by using both of the time thresholds. That means for each candidate session constructed, both the total duration time of session and the time spent on a page in a session will be limited. 

\textbf{Phase 2:} This phase corresponds to MPVS problem and constructs all maximal navigation sequences from the candidate sessions generated at the first phase. The following features related to the navigation sequences are used in this phase:

\begin{itemize}
	\item \textbf{Maximality:} During the execution of phase two, each new sequence which is either constructed by adding a page to an existing sequence or constructed from a single page, is maximal at the beginning. A sequence becomes non-maximal if a new navigation sequence is constructed from it by adding a page to its tail.

 \item \textbf{Degree:} The degree of a sequence indicates how many new sequences can be constructed from it by adding new pages to its tail. Thus, the degree of a sequence is equal to the out-degree of its last page when it is constructed. With the extension by appending a new page, the degree of the current navigation sequence is decreased by one which also makes the extended sequence non-maximal. Moreover, non-maximal sequences must be kept as long as they are extendable, i.e., their degrees are still larger than zero.
\end{itemize}

\begin{algorithm}
\caption{CSRA}
\begin{algorithmic}[1]
\STATE \textbf{input:} CandidateSessionSet
	\STATE \textbf{output:} MSeqSet
	\STATE  \textbf{global variables:} FSeqSet, TSeqSet 
	\STATE  \textbf{global variables:} MSeqSet, flag 
		\STATE \textbf{procedure} \underline{MPVS} $(\texttt{CandidateSession})$
    \STATE \texttt{ }  \textbf{for each}  $\texttt{WP}_{\texttt{i}}$  \textbf{in} CandidateSession  $//$ $\texttt{WP}_{\texttt{i}}$ is i-th web page.
    \STATE \texttt{ }\texttt{ } flag := FALSE
    \STATE \texttt{ }\texttt{ } \textbf{for each} $\texttt{Seq}_{\texttt{j}}$ \textbf{in} TSeqSet
    \STATE \texttt{ }\texttt{ }\texttt{ } \textbf{newSeqExtend}($\texttt{Seq}_{\texttt{j}}$, $\texttt{WP}_{\texttt{i}}$)
    \STATE \texttt{ }\texttt{ } \textbf{end for each}
    \STATE \texttt{ }\texttt{ }  \textbf{if} flag = FALSE \textbf{then}   
    \STATE \texttt{ }\texttt{ }\texttt{ }\texttt{ } \textbf{newSeqInitialize}($\texttt{WP}_{\texttt{i}}$)
    \STATE \texttt{ }\texttt{ }  \textbf{end if}
    \STATE \texttt{ }\texttt{ }\textbf{end for each}
    \STATE \textbf{end procedure}

		\STATE \textbf{procedure} \underline{CSRA\_Phase\_2} $(\texttt{CandidateSessionSet})$
    \STATE \texttt{ } MSeqSet := \{\} //Maximal Sequences
    \STATE \texttt{ } \textbf{for each} CandidateSession \textbf{in} CandidateSessionSet 
    \STATE \texttt{ }\texttt{ } TSeqSet := \{\}  //Temporary Sequences
    \STATE \texttt{ }\texttt{ } FSeqSet := \{\}  //Final Sequences
    \STATE \texttt{ }\texttt{ } \textbf{MPVS} (CandidateSession) 
    \STATE \texttt{ }\texttt{ } \textbf{for each} $\texttt{Seq}_{\texttt{j}}$ \textbf{in} TSeqSet 
    \STATE \texttt{ }\texttt{ }\texttt{ } \textbf{if} $\texttt{Seq}_{\texttt{j}}$.maximal = TRUE \textbf{then}
    \STATE \texttt{ }\texttt{ }\texttt{ }\texttt{ } FSeqSet := FSeqSet $\cup$ {$\texttt{Seq}_{\texttt{j}}$}
    \STATE \texttt{ }\texttt{ }\texttt{ } \textbf{end if}
    \STATE \texttt{ }\texttt{ }\textbf{end for each}
    \STATE \texttt{ }\texttt{ } MSeqSet := MSeqSet $\cup$ FSeqSet
    \STATE \texttt{ }\textbf{end for each}
    \STATE \textbf{end procedure}
		
\end{algorithmic}
\end{algorithm}

The following global variables are used in the second phase of C-SRA:
\begin{itemize} 
\item \textbf{FSeqSet:} Maximal navigation sequences with degrees zero generated from current candidate sessions.
\item \textbf{TSeqSet:} Navigation sequences with degrees greater than zero, it can still contain maximal sessions.
\item \textbf{MSeqSet:} Collection of all maximal navigation sequences obtained from all candidate sessions. 
\end{itemize}

The details of the second phase of C-SRA are given in Algorithm-1 and Algorithm-2 respectively. In the main procedure, called CSRA\_Phase\_2, each candidate session of the CandidateSessionSet is processed by calling the procedure MPVS. In this procedure each page in the candidate session is processed from left to right to determine if that page can expand existing navigation sequences or it can initiate a new sequence. At any particular step, if the degree of any maximal sequence decreases to zero, it is automatically added to final sequence set since there is no page remained in the candidate session to expand current sequence. After completing the processing of each page of a candidate session, maximal sequences remaining in the temporary sequences set with out-degrees greater than zero are also added to the final sequence set. Finally, at the end of processing each candidate sessions, the final sequence set obtained from that candidate session is added to the global maximal sequences set.

\begin{figure}[h!]
	\centering
		\epsfig{file=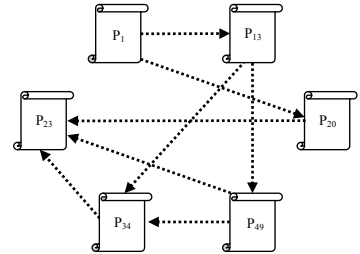, width=0.55\linewidth}
	\caption{\small An example web site topology graph \label{fig:deneme}}
\end{figure}

Table~\ref{tab:exC-SRA} illustrates the execution of the Phase 2 of C-SRA for the candidate session $[P_{1}, P_{20}, P_{23}, P_{13}, P_{34}]$ corresponding to the site topology in Figure~\ref{fig:deneme}. In the table, each column represents the processing single page of the candidate session and navigation sequences are shown together with their degrees and maximality flags (as triples of $<$sequence: degree: maximality flag$>$). 

\begin{algorithm}[h!]
\caption{Session Functions}
\begin{algorithmic}[1]
	\STATE \textbf{input:} CandidateSessionSet
	\STATE \textbf{output:} MSeqSet
	\STATE  \textbf{global variables:} FSeqSet, TSeqSet 
	\STATE  \textbf{global variables:} MSeqSet, flag 
		\STATE \textbf{procedure} \underline{NewSeqExtend} $(\texttt{Seq}_{\texttt{j}}, \texttt{WP}_{\texttt{i}})$    
    \STATE \texttt{ } linkStatus := Link[LastElement($\texttt{Seq}_{\texttt{j}}$), $\texttt{WP}_{\texttt{i}}$]
    \STATE \texttt{ } timeDiff\texttt{ }\texttt{ }:= \textbf{TimeDiff}(LastElement($\texttt{Seq}_{\texttt{j}}$),$\texttt{WP}_{\texttt{i}}$)
    \STATE \texttt{ } \textbf{if} linkStatus=true \textbf{and} timeDiff $<$ $\delta$ \textbf{then}
			\STATE \hspace{6 mm}\texttt{flag} := TRUE
    	\STATE \hspace{6 mm}$\texttt{Seq}_{\texttt{j}}$.degree := $\texttt{Seq}_{\texttt{j}}$.degree -1
   		\STATE \hspace{6 mm}$\texttt{Seq}_{\texttt{j}}$.maximal := FALSE
			\STATE \hspace{6 mm}\texttt{NewSeq}.degree := $\texttt{WP}_{\texttt{i}}$.outdegree 
			\STATE \hspace{6 mm}\texttt{NewSeq}.maximal := TRUE 
			\STATE \hspace{6 mm}\texttt{NewSeq} := $\texttt{Seq}_{\texttt{j}}$ $\bullet$ $\texttt{WP}_{\texttt{i}}$   //Append
			
			 \STATE \hspace{6 mm}\textbf{if} NewSeq.degree = 0 \textbf{then}
    	 \STATE \hspace{8 mm}FSeqSet := FSeqSet U \{NewSeq\}
       \STATE \hspace{6 mm}\textbf{else}  
       \STATE \hspace{8 mm}TSeqSet := TSeqSet U \{NewSeq\}
       \STATE \hspace{6 mm}\textbf{end if}

			 \STATE \hspace{6 mm}\textbf{if} $\texttt{Seq}_{\texttt{j}}$.degree = 0 \textbf{then}
    	 \STATE \hspace{8 mm}TSeqSet := TSeqSet - \{$\texttt{Seq}_{\texttt{j}}$\}
       \STATE \hspace{6 mm}\textbf{end if}
		\STATE \texttt{ } \textbf{end if}
		\STATE \textbf{end procedure}

		\STATE \textbf{procedure} \underline{NewSeqInitialize} $(\texttt{WP}_{\texttt{i}})$
		
    \STATE \texttt{ } NewSeq.degree := $\texttt{WP}_{\texttt{i}}$.outdegree
    \STATE \texttt{ } NewSeq.maximal := TRUE
    \STATE \texttt{ } NewSeq :=  [$\texttt{WP}_{\texttt{i}}$]
    \STATE \texttt{ } \textbf{if} NewSeq.degree = 0 \textbf{then}
    \STATE \hspace{6 mm} FSeqSet := FSeqSet U \{NewSeq\}
    \STATE \texttt{ } \textbf{else } 
    \STATE \hspace{6 mm} TSeqSet := TSeqSet U \{NewSeq\}
    \STATE \textbf{end procedure}
\end{algorithmic}
\end{algorithm}

Notice that, in this example we assume that there is no case that violates page stay time constraint. Referring to Table 1, when page $P_{1}$ is processed, since it is the first page, a new sequence containing only page $P_{1}$ is created. Since it is a new page, it is maximal (T represent the maximality is true), and its out-degree is the out-degree of the page $P_{1}$, which is 2. When the second page, $P_{20}$, is processed, the sequence $[P_{1}]$ will be extended and a new sequence $[P_{1}, P_{20}]$ will be generated. The new Sequence $[P_{1}, P_{20}]$ will be maximal, but, we will still keep the extended Sequence $[P_{1}]$ in the temporary sequence set. Although its out-degree is decreased, since it is still greater than zero (which is 1) it can still be extended by using its unused hyperlinks (out-going edges in the graph representation). Moreover, since the sequence $[P_{1}]$ was extended, its out-degree was decreased to 1 and it becomes non-maximal. The new sequence, $[P_{1}, P_{20}]$, on the other hand is marked as maximal with the out-degree 1, which is the out-degree of its last page $P_{20}$. After that, the new sequence, $[P_{1}, P_{20}, P_{23}]$ is obtained while processing the third page, $P_{23}$.  This sequence has an out-degree 0, thus rather than keeping it in the temporary Sequence set, it is directly moved into the final Sequence set, since it can not be expanded any further. After processing the last two pages of the candidate session ($P_{13}$ and $P_{34}$), the sequence $[P_{1}, P_{13}, P_{34}]$ was also generated. Since this sequence has a degree 1, it is placed into the temporary Sequence set. After all pages in the candidate session have been processed, this sequence is also moved into the Final Sequence set due to maximality. As a result, after completing the processing of the candidate session $[P_{1}, P_{20}, P_{23}, P_{13}, P_{34}]$, the second phase of C-SRA discovers two maximal sequences: $[P_{1}, P_{20}, P_{23}]$ and $[P_{1}, P_{13}, P_{34}]$.

\begin{table}[h!]
\centering
\caption{Execution of Complete-SRA \label{tab:exC-SRA}}
\begin{tabular}{|c|c|c|} \hline
\textbf{\tiny \textbf{Page}}& {\tiny \textbf{$P_{1}$}} & {\tiny \textbf{$P_{20}$}}\\ \hline 
{\tiny Temp Sequences} & & {\tiny $<[P_{1}]:2:T>$} \\ \hline
{\tiny Extended Set} & & {\tiny $<[P_{1}]:1:F>$} \\ \hline
{\tiny New Sequence} & {\tiny $<[P_{1}]:2:T>$} & {\tiny $<[P_{1}, P_{20}]:1:T>$} \\ \hline
{\tiny Final Set} & & {}\\ \hline

\textbf{\tiny \textbf{Page}}& { \tiny \textbf{$P_{23}$}} & {\tiny \textbf{$P_{13}$}}  \\ \hline 
\multirow{2}{*}{\tiny Temp Sequences} & {\tiny $<[P_{1}]:1:F>$} & { \tiny $<[P_{1}]:1:F>$} \\
																		 & {\tiny $<[P_{1}, P_{20}]:1:T>$} & \\ \hline
{\tiny Extended Set} & {\tiny $<[P_{1}, P_{20}]:0:F>$} & {\tiny $<[P_{1}]:0:F>$} \\ \hline
{\tiny New Sequence} & {\tiny $<[P_{1}, P_{20}, P_{23}]:0:T>$} & {\tiny $<[P_{1}, P_{13}]:2:T>$} \\ \hline
{\tiny Final Set} & {\tiny $<[P_{1}, P_{20}, P_{23}]:0:T>$} & {\tiny $<[P_{1}, P_{20}, P_{23}]:0:T>$}\\ \hline

\textbf{\tiny \textbf{Page}}& \multicolumn{2}{c|}{\tiny \textbf{$P_{34}$}} \\ \hline 
{\tiny Temp Sequences}& \multicolumn{2}{c|}{\tiny $<[P_{1}, P_{13}]:2:T>$} \\ \hline
{\tiny Extended Set}& \multicolumn{2}{c|}{\tiny $<[P_{1}, P_{13}]:1:F>$} \\ \hline
{\tiny New Sequence}& \multicolumn{2}{c|}{\tiny $<[P_{1}, P_{13}, P_{34}]:1:T>$} \\ \hline
\multirow{2}{*} {\tiny Final Set}& \multicolumn{2}{c|}{\tiny $<[P_{1}, P_{20}, P_{23}]:0:T>$} \\
&  \multicolumn{2}{c|}{\tiny $<[P_{1}, P_{13}, P_{34}]:1:T>$} \\ \hline

\end{tabular}
\end{table}

\section{Experiments and Discussions}

Sequential pattern discovery is the next phase of the Web Usage Mining after session construction. In this phase, frequent user access patterns are discovered from session sequences. In our experiments, we applied our web usage mining component (C-SRA + Pattern Discovery) to the server logs generated by simulator described in~\cite{Bayir09}. We have compared our algorithm by replacing C-SRA in the WUM tool with previous session construction techniques~\cite{Liu11}(time and navigation oriented techniques) in terms of accuracy metric introduced in~\cite{Bayir09}. Our experiments show that C-SRA performs at least 20\% - 25\% better than previous approaches.

%
\bibliographystyle{abbrv}

\begin{thebibliography}{1}

\bibitem{Bayir09}
M.~A. Bayir, I.~H. Toroslu, A.~Cosar, and G.~Fidan.
\newblock Smart miner: a new framework for mining large scale web usage data.
\newblock In {\em WWW}, pages 161--170, 2009.

\bibitem{Bayir12}
M.~A. Bayir, I.~H. Toroslu, M.~Demirbas, and A.~Cosar.
\newblock Discovering better navigation sequences for the session construction
  problem.
\newblock {\em Data Knowl. Eng.}, 73:58--72, 2012.

\bibitem{Liu11}
B.~Liu, B.~Mobasher, and O.~Nasraoui.
\newblock Web usage mining.
\newblock In {\em Web Data Mining}, Data-Centric Systems and Applications,
  pages 527--603. Springer Berlin Heidelberg, 2011.

\bibitem{Mobasher07}
B.~Mobasher.
\newblock Data mining for web personalization.
\newblock In {\em The Adaptive Web}, pages 90--135, 2007.

\end{thebibliography}

\end{document}